\documentclass[twocolumn,showpacs,preprintnumbers,amsmath,amssymb]{revtex4}
\usepackage{amssymb}
\usepackage{graphicx}
\usepackage{natbib}
\usepackage{epstopdf}
\usepackage{dcolumn}
\usepackage{bm}
\usepackage{makeidx}
\usepackage{enumerate}
\begin{document}

\title{Branching Transport Model of Alkali-Halide Scintillators}

\author{B. S. Alexandrov},
 \affiliation{Nuclear Nonproliferation Division, Los Alamos National Laboratory, Los Alamos, NM 87545, USA, } 
 \affiliation{the University of New Mexico, Albuquerque, NM 87131, USA}
  
\author{K. D. Ianakiev}%
\affiliation{Nuclear Nonproliferation Division, Los Alamos National Laboratory, Los Alamos, NM 87545, USA} 
\author{P. B. Littlewood}
\affiliation{Cavendish Laboratory, Cambridge University, Cambridge CB3 0HE, UK
}%

\date{\today}

\begin{abstract}

We measure the time dependence of the scintillator light-emission pulses in NaI(Tl) crystals at different temperatures,
 after activation by gamma rays. We confirm that there are two main nonexponential components to the time decay and find that their amplitude ratio shows Arrhenius temperature dependence. We explain these nonexponential components as arising from two competing mechanisms of carrier transport to the Tl activation levels. The total light output of the NaI(Tl) detectors shows a linear temperature dependence explained by our model. 

\end{abstract}
\maketitle

\section{Introduction}
Doped alkali-halide scintillation crystals, developed 50 years ago, continue to be the workhorses of outdoor gamma spectroscopy because of their size, cost, and performance. Recently, there has been a growing interest in NaI(Tl) detectors for use in Homeland Security applications  \cite{1, 2}. However, a well-known, problematic feature of NaI(Tl) detectors is the temperature-dependence of their light yield. The core of this problem stems from the temperature dependence of the shape and amplitude of the light pulses emitted from the scintillator for a given energy of the incident ionizing particle. Although multiple exponential components describing the shape of the light pulse in time have been reported  \cite{3}, the common understanding for a NaI(Tl) light pulse assumes a single dominant exponential component with a temperature-dependent decay-time constant \cite{4,5,6,7,8}, based on the effective Òthree-level activator modelÓ described in  \cite{23}. This is inconsistent with the well known fact that below 60$^0$C, the current pulses cannot be fitted well with a single exponent, even at a  fixed temperature.  Until now, it has been also generally accepted that the temperature behavior of the total light output of NaI(Tl) is nonlinear and has a broad maximum below room temperature.\cite{3,4,5,6}.

Recent experimental results obtained by using a novel method and device  demonstrated that this experimentally observed nonlinear temperature behavior arises because of a convolution between the temperature-dependent shape of the light pulse and the pulse response of the shaping circuitry \cite{9}. It was shown experimentally that when a gated integrator is used, the whole light output shows only linear temperature dependence over a wide temperature range (-30$^0$C to +60$^0$C) \cite{9,10,11,12}. In the same work, two main components of the shape of the light pulses, with a temperature redistribution between their amplitudes, were found. Importantly, it was shown in  \cite{10,11,12} that the slow component is negligible above room temperature, but it produces up to 40 percent of the total light at $ -20^0$C and lasts several microseconds. It was also shown that at high temperatures, only one almost exponential decay component exists, consistent with  \cite{5}.

We measured the time dependence of NaI(Tl) current pulses at different temperatures. We show that rather than one or more exponential decay components, there are two dominant nonexponential light components, corresponding to two competing mechanisms of carrier transport to the Tl activation levels. We model the transport and reproduce the experimental data with nonlinear rate equations. The data may also be fitted with two exponential decay components using fixed fast and slow time constants and an Arrhenius temperature-dependent redistribution between the amplitudes, but the accuracy is  approximately 5 percent.

\section{EXPERIMENTAL RESULTS}
We placed a standard 2-in x 2-in.  Bicron NaI(Tl) detector in an environmental chamber. The chamber temperature changed at a rate of 2$^0$C/hr, and each set of measurements at a given temperature was taken after an 8-hr hold time to allow good thermal equilibration. We digitized the photomultiplier tube (PMT) current pulses with a  14-bit digital scope, manufactured by Gage Applied Sciences Inc.,  connected via a $50$ $\Omega$   resistor directly to the anode of the PMT for each temperature. The temperature coefficient of the PMT was almost constant over the investigated temperature interval \cite{24} and did not change the shape of the light pulse. The area of the pulses was selected to correspond to an excitation around 662 keV. We normalized all pulses to the unit area to study the effect on pulse shape alone. In Fig. 1, we show the time dependence of the emitted light output on a  log scale, where the two components are resolved and the temperature-dependent redistribution between the amplitudes of the slow and fast components is readily seen. Note that the emission spectrum of this doped inorganic scintillator shows that the emitted light is at a single wavelength, characteristic of a transition between a single excited state of the activator and its ground state. Thus the two components of the light pulse do not represent decays from different energy levels, but rather, they represent two different pathways for the secondary electrons and holes to the activator levels.
\begin{figure}[!h]
\begin{center}
\includegraphics[width=\columnwidth]{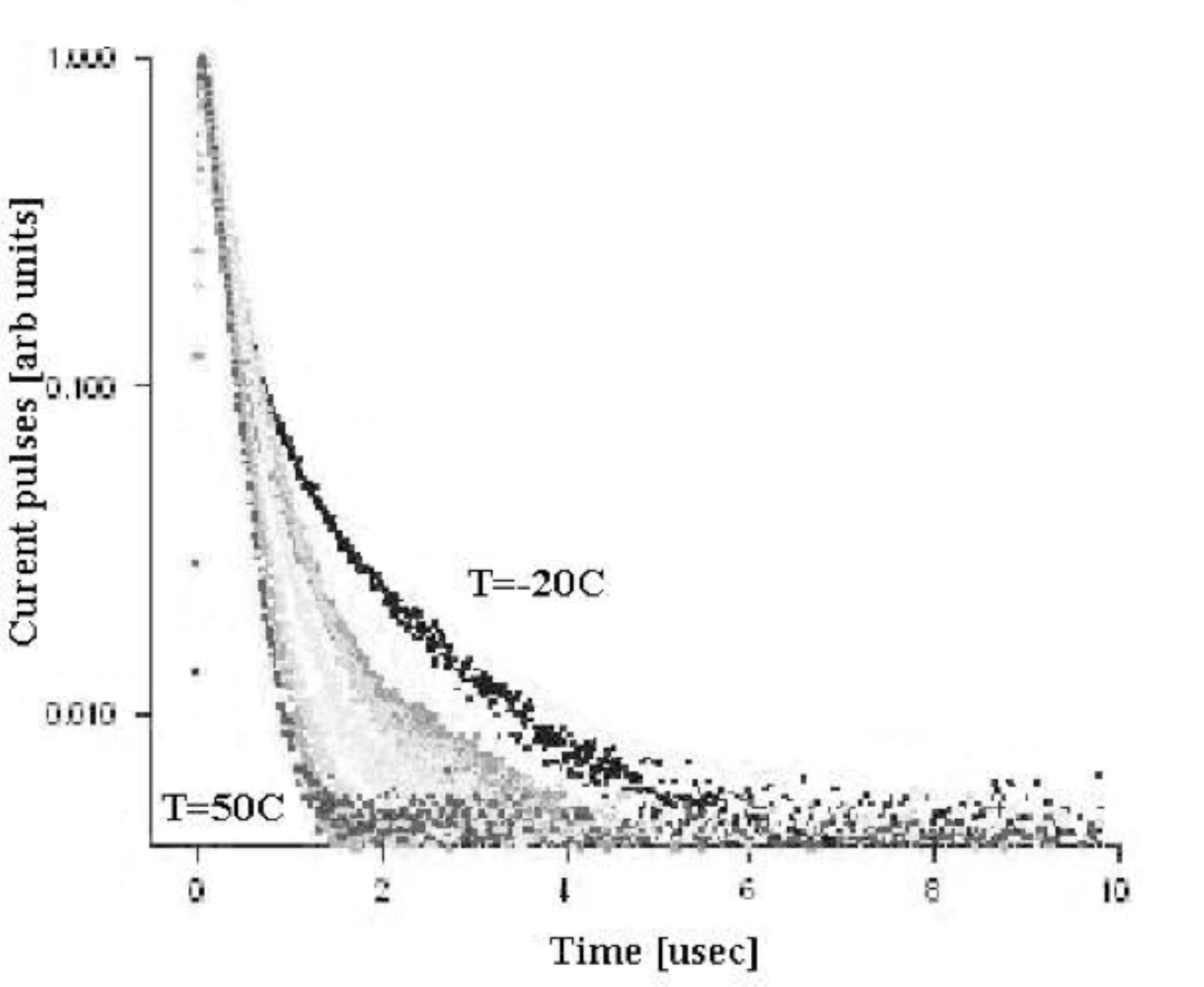}
\end{center}   
\caption{Current pulses measured at $-20^0C, -10^0C,0^0C,10^0C,20^0C,30^0C,50^0C$. Each one normalized to its maximum value so the second component may be clearly resolved at lower temperatures.}
\label{fig:1}
\end{figure} 
 We normalized the area of the current pulses  to remove the temperature-dependence arising from a competition between the scintillation transitions and the nonradiative phonon-assisted transitions. This competition take place at the excited Tl level, where the electrons and holes recombine \cite{3}. The nonradiative, phonon-assisted transitions decrease the total amount of the collected light, but they play this role after the secondary carriers, electrons and holes, are recombined at the (Tl$^+$)$^*$ level. The nonradiative, phonon-assisted transitions do not significantly change  the time shape of the current pulse. This can readily  be understood  at very higher temperatures where  only one exponential component of the current pulses exists \cite{5} and the time shape of the pulse is almost temperature independent. On the other hand the nonradiative, phonon-assisted transitions are responsible for  the strong temperature dependence of the peak position, and this dependence can  be observed at arbitrarily higher  temperatures. At such high temperatures,  the shape of the current pulse does not change with temperature, but the area of the pulse and the corresponding peak position for a given energy of the incident particle becomes less and less with increasing temperature.. 

\begin{figure}[!h]
\begin{center}
\includegraphics[width=\columnwidth]{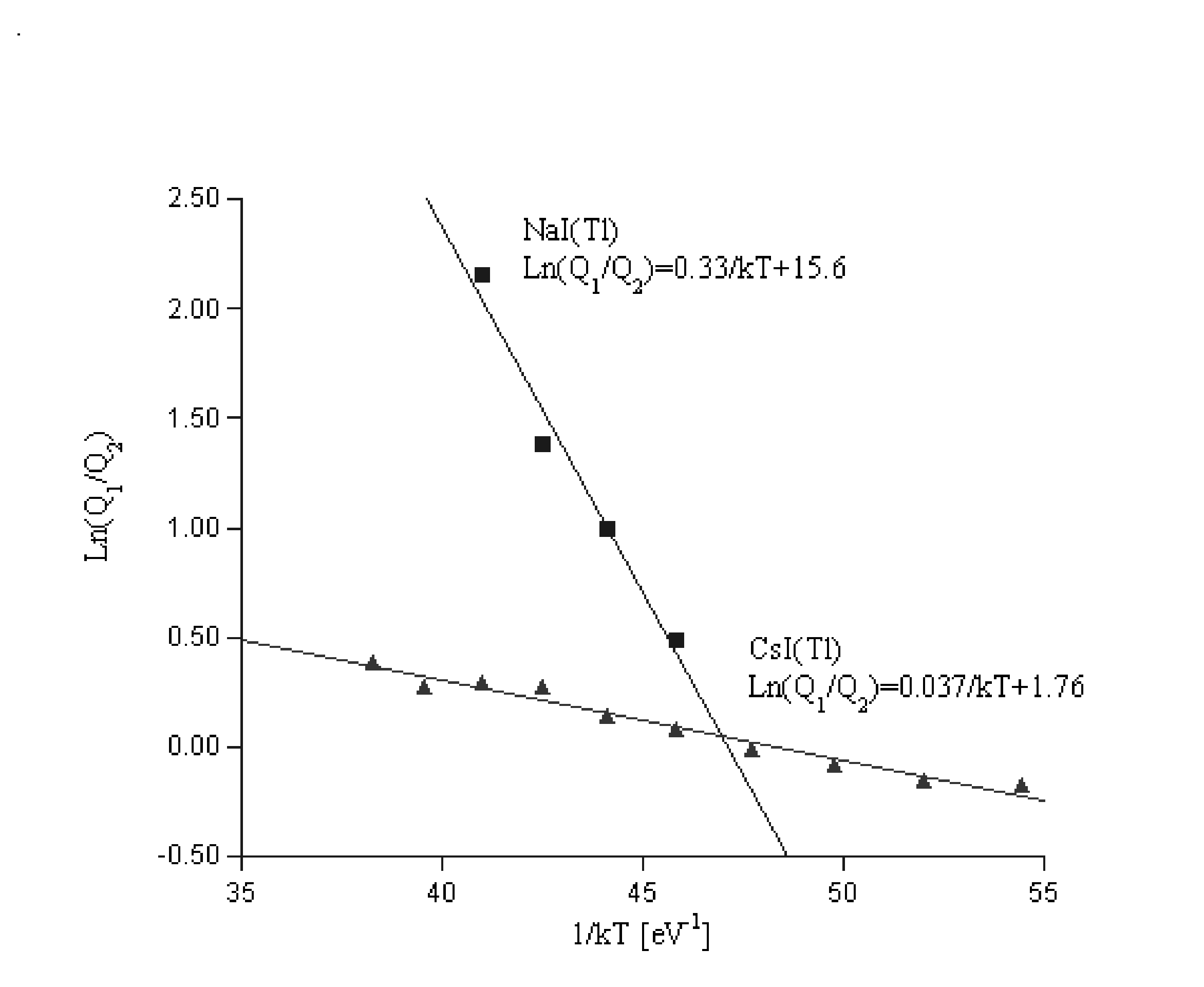}
\end{center}   
\caption{Activation dependence of the total amplitude of the current pulses for NaI(Tl) (squares) and CsI (Tl) (triangles), on a logarithmic scale. Numerical values for the amplitudes Q$_1$ and Q$_2$ of CsI(Tl) by courtesy of \cite{22}.}
\label{fig:2}
\end{figure}

In our measurements of the light pulses from NaI(Tl), we found an Arrhenius dependence of the ratio between the amplitudes of the two dominant components with which we fit the data. We found a similar dependence in CsI(Tl) using the numerical data published in \cite{22}, as shown in  Fig. \ref{fig:2}. We found that
\begin{eqnarray}
 \frac{Q_1}{Q_2}\propto e^{(-\frac{\Delta E}{kT})},
 \label{eq:1}
\end{eqnarray}
 where $Q_1(T)$ and $Q_2(T)$ are the amplitudes of the fast and the slow components, $Q_1(T) + Q_2(T) = constant$, and $\Delta E$ is a phenomenological activation energy between the STE level and  the excited Tl level.

\section{THEORETICAL MODEL}
Because we were sure that the observed components of the light pulses did not correspond to impurities or different energy levels, we sought to determine the primary transport mechanisms for secondary electrons and holes.

In alkali-halide crystals, the electron-hole interaction is very strong \cite{13}. This leads to the formation of excitons and, because of the  highly polarizable nature of the crystals, to a fast creation of self-trapped excitons (STEs) at the very beginning of the process. This is why the earliest theoretical models of the transport in NaI(Tl) included only the motion of the excitons \cite{14}. There is considerable evidence of  long-lived  STEs in alkali-halide crystals at low temperatures \cite{19}. In  NaI(Tl),  a redistribution between the emitted light from the STE levels and the emitted light from the Tl activation centers   above T= 140 K has been observed. This means that after the decay of an STE, the created electron and hole usually reach the activator centers \cite{20}. The suggested transport mechanism is by phonon-assisted exciton hopping from the STE level to the Tl level. In NaI(Tl), this type of transport was observed  with a temperature-dependent rate  of $10^{12}-10^{8} [s^{-1}]$ \cite {20}. 

On the other hand, a previously performed experiment on CsI(Tl) showed that the predicted dependence of the shape of the light pulse on the activator concentration, for different energies of an  incident particle, is not observable \cite{14,15}. This disagreement with the existing  theory led to the incorporation of a second kind of motion in the crystal: namely, the binary diffusion of electrons and holes \cite{16}. The binary diffusion in these materials leads to the fast creation of localized electrons and holes in the vicinity of the activator centers and, after that, to complex mechanisms of recombination. Optical measurements \cite{17,18} proved that the lifetime of holes trapped at $Tl^{++}$ is surprisingly short: approximately 0.350 $\mu$sec in NaI(Tl). Therefore, we considered another possible mechanism of STE decay, leading to the creation of separated electrons and holes and to the diffusion-transport mechanism. The mechanism we considered is the previously discovered multiphonon-assisted dissociation of the STE \cite{21}. The corresponding rate is again temperature dependent, and its value is in the same range as the rate of the hopping transport.

\begin{figure}[!h]
\begin{center}
\includegraphics[width=\columnwidth]{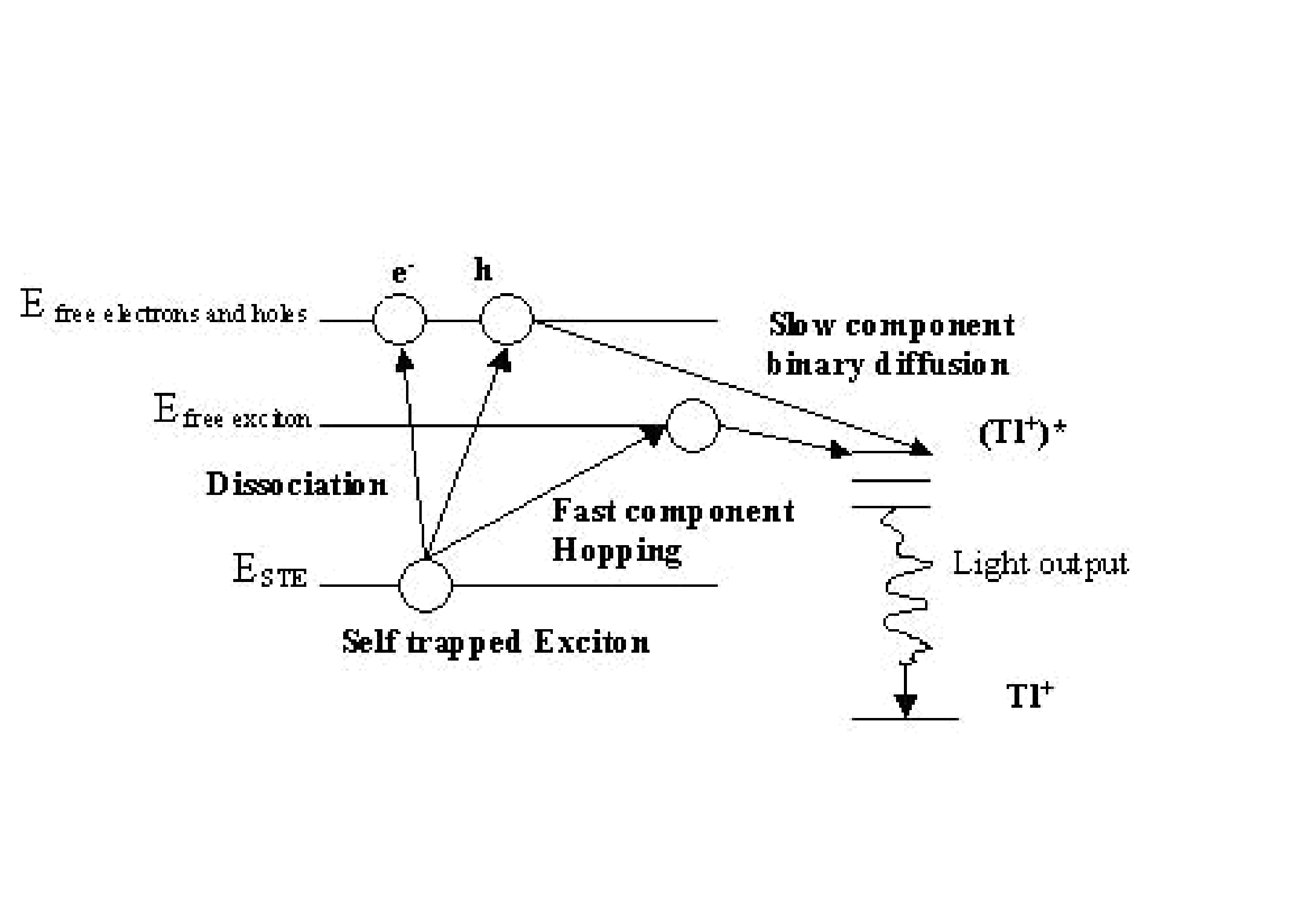}
\end{center}
\caption{Diagram representing  the two main competing processes in alkali-halide crystals. Intermediate processes creating $Tl^{++}$ and $Tl^0$ are omitted. The quenching levels in $Tl^+$ are not shown because the nonradiative transitions change only the integrated light output. They are responsible for the linear temperature dependence of the total light output, but they do not change the temporal shape of the light emission. }
\label{fig:3}
\end{figure}

  The simple relationship  shown in Eq.(\ref{eq:1})  suggests that there are two transport mechanisms in competition for reaching the activated site. We propose that one is a phonon-assisted hopping transport of the STE to the Tl level, and the other is a multiphonon-assisted dissociation of the STE followed by single-carrier transport of the electron and hole to the Tl level, as shown in Fig \ref{fig:3}.  In the first channel, the STE reaches the activator levels $(Tl^+)$ via phonon-assisted hopping, making $(Tl^{+})^*$ excited levels that creates the fast component of the light output. The decay time of this component would be mainly the lifetime of the $(Tl^{+})^*$ excited levels. We can infer that the combination of the lifetime of the STE level and the STE transport to the activation center is relatively fast. Note that the decay rate is determined also by the nonradiative, phonon-assisted transition between  $(Tl^{+})^*$  and the ground state $(Tl^{+})$.
  
We associate the second (slower) component with binary transport  \cite{17}.The mechanism is by STEs that thermally dissociate into electrons and holes. The electrons are rapidly captured at  $(Tl^{+})$  levels, making  $(Tl^{0})$  levels, and the holes are quickly captured at   $(Tl^{+})$  levels, making  $(Tl^{++})$  levels. The holes may reach the  $(Tl^{0})$  levels via diffusion and recombine with the electrons, thus creating  $(Tl^{+})^*$  excited levels that decay optically, as before. The lifetime of this process is slower than the hopping transport   because it is a combination of the relatively long lifetime of the trapped hole, the diffusion time, and the lifetime of the  $(Tl^{+})^*$  excited level. The two processes are schematically depicted in Fig. \ref{fig:3}.

Generally speaking, it is clear that both parts of the response are not simply exponential, being a convolution of the lifetimes of the trapped hole, diffusive transport, and the $(Tl^{+})^{*}$ level in the slow component, and of the lifetime of the $STE$ level, hopping transport, and lifetime of the $(Tl^{+})^{*}$  level in the fast component. 

We now compare this physical picture with the traditional  one that is currently used   \cite{3,4,5}.  Making an approximation, the established  model is based on a single temperature-dependent decay-time constant \cite{5}. This model actually describes the lifetime of the activator level \cite{3}. It is supported from a  simple consideration  of the activator's center consisting of three energy levels: the ground level G, the scintillation level S, and the quenching level Q \cite { 23}. The calculations of the quantum efficiency q in this three levels-model  lead to

\begin{eqnarray} 
&q&=\frac{ k_S\exp{ (-\frac{W_S}{kT}) }} {k_S\exp{(-\frac{W_S}{kT})}+k_Q\exp{(-\frac{W_S+W_Q}{kT}})}\nonumber, \\
&q&=\frac{ 1} {1+\frac{k_Q}{k_S}\exp{(-\frac{W_Q}{kT}})}.
\label{eq:2} 
\end{eqnarray}

Three levels model takes into account that at  temperature T, the relative populations of the levels are as follows: level S $ \sim \exp {(-\frac{W_S}{kT})}$; level Q  $ \sim \exp {(-\frac{W_S+W_Q}{kT})}$. Here $ k_S$ is the probability of  radiative transition from  radiative level S  to the ground level G, and $k_Q$ is the probability for nonradiative transition from the quenching level Q to the ground level (so called internal quenching). $W_S$ is  the energy difference between scintillation level S and ground level G, and  $W_Q$ is the energy difference between quenching level Q and scintillation level S.  The  quantum efficiency calculated in this way is a monotonically increasing function with decreasing temperature, and one would expect a monotonic temperature dependence of the total light output.

However, experimentally, the temperature dependence of the light output from the NaI(Tl) scintillator was found to be strongly nonlinear \cite{3,4}, possessing a broad maximum below room temperature,  in disagreement with the monotonic behavior of the quantum efficiency of the model. An  additional function $f(T)$ (so called external quenching)  that represents the probability of the occupation  of  the activator's centers or, in other words, describes the temperature dependence of total charge $ Q_{total}=Q_{total}(T)$, was introduced to multiply the quantum efficiency in order to explain this nonlinear behavior \cite{3,4}. This function has been introduced to describe the nonlinear shape of the temperature dependence of the total light output. But as we pointed out, the nonlinear temperature dependence does not exist if one collects the total light using a gated integrator \cite{9}. The nonlinear behavior of the peak position with temperature is different for different shapers \cite{9}, because the long component produces a different fraction of the whole light at different temperatures  \cite{9,10,11,12}. Using different shapers, we do not gather the whole light, but we cut off part of the signal produced by the long component, and therefore, we cut off part of the light that was produced via the slower pathway to the Tl level. This cutting is temperature dependent because the amount of the light that we cut belongs to the long component, and  the  amount of the light belonging to the long component is temperature dependent \cite{9,10,11,12}. The existing models for the temperature behavior of the NaI(Tl) do not take into account the complex shape of the individual light pulse, connected with the redistribution between the slow and fast component, or the simple linear temperature dependence of total light output that was recently measured \cite{9}.

In our model, the total charge created from the incident ionizing particle  has two main components. As we discussed above, we found experimentally that their amplitude ratio shows an Arrhenius temperature dependence, whereas their sum is a constant. This means that
\begin{eqnarray*}
 &Q_{hopping}(T)&= \frac{Q_{total}A \exp{\frac{\Delta E}{kT}}}{1+A\exp{\frac{\Delta E}{kT}}},\\ 
 &Q_{dissociation}(T)&=\frac{Q_{total}}{1+A\exp{\frac{\Delta E}{kT}}}.\\
 \end{eqnarray*} 
 
Here, $ Q_{hopping}(T)$ and $ Q_{dissociation}(T)$  correspond to charges that use two different pathways for reaching the activator levels, and $A$ is the ratio between the probability for STE decay via multiphonon dissociation and the probability for STE decay via hopping at  very high  temperature. Both components of the total charge are strongly temperature dependent, but their sum does not depend on temperature. That is why in our model  we have $f(T)=1$. As a consequence, we have temperature dependence of the total light output, as in the pure  three-level activator model. The  quantum efficiency calculated in this model, pointed out in  Eq.(\ref{eq:2}), is  a monotonically increasing function with decreasing temperature. This means that in a narrow temperature interval (kT in the range 0.03 to 0.02 eV), it can be approximated with  a  linear temperature dependence. Therefore in such a narrow temperature interval (T in the range -$30^0$C to +$60^0$C ),  the total collected charge $ Q_{total}$  depends almost linearly on temperature, and because of that, we will observe an almost  linear temperature dependence of the total light output $L(T)$: 
\begin{eqnarray*} 
L(T)\simeq -CT+1,
\end{eqnarray*}
where $C$ is a constant, and we measure in percent $L(T)$ as it was shown in \cite{9,10,11,12}.  At the same time,  we have a strong temperature dependence and redistribution between the two main components of  the light pulse, which explains the variety of the nonlinear temperature dependences of the light output when different shapers are used \cite{9}.
  
To check the model,  we  simulated  the processes considered above using rate equations that describe the populations of the STEs, separated electrons and holes, populations of the  $(Tl^{++})$, $(Tl^0)$ levels, and the final population of the $(Tl^{+})^*$  level, which gave us the time dependence of the light output: 
\begin{eqnarray*}
\frac{dSTE}{dt}& = & - \frac{  STE} {  \tau_{hop} } -      \frac{ STE} {  \tau_{ diss} } ,\\
\frac{d{e^{-} }}{dt}& = &  \frac{ STE} { \tau_{diss} }-\frac{ {e^{-} } } {  \tau^{0}_{trap} }+\frac { Tl^{0}_{e^{-} } }{  \tau^{0}_{dtrap} }-\frac{ {e^{-} }  Tl^{++}_{h^{+} } }{  \tau_{aftgl} },\\
\frac{d{h^{+}}}{dt}& = & \frac{ STE}{  \tau_{diss} }-\frac{ {h^{+} } }{  \tau^{++}_{trap} }+\frac{Tl^{++}_{h^{+} } }{\tau^{++}_{dtrap} }-\frac{{h^{+} } Tl^{0}_{e^{-} } }{\tau_{diff} },\\
\frac{dTl^{0}_{e^{-}}}{dt}& = & \frac{ {e^{-}} }{ \tau^{0}_{trap} }-\frac{ Tl^{0}_{e^{-} } }{ \tau^{0}_{dtrap} }-\frac{{h^{+}} Tl^{0}_{e^{-} } }{\tau_{diff} },\\
\frac{dTl^{++}_{h^{+}}}{dt}&= & \frac{{h^{+}}}{\tau^{++}_{trap}}-\frac{Tl^{++}_{h^{+}}}{\tau^{++}_{dtrap}}-\frac{{e^{-}} Tl^{++}_{h^{+}}}{\tau_{aftgl}},\\
\frac{dTl^{+*}_{popul}}{dt}&=  &\frac{STE}{\tau_{hop}}+\frac{{h^{+}} Tl^{0}_{e^{-}}}{\tau_{diff}}+\frac{{e^{-}} Tl^{++}_{h^{+}}}{\tau_{aftgl}}-\frac{Tl^{+*}_{popul}}{\tau_{scin}}.
\end{eqnarray*}

The equations include nonlinear bimolecular terms describing the diffusion of the holes from $(Tl^{++})$ to $(Tl^{0})$ and thermoactivated transport of the electrons from $(Tl^0)$ to $(Tl^{++})$ that are responsible for afterglow. We  numerically solved this stiff system of ordinary differential equations using a standard MatLab program, $\it{ode15s}$. \\
Our variables are as follows:
\begin{enumerate}[(1)]
\item  $STE$  -    density of the  STEs.
\item $e^{-}$  -  density of  the electrons.
\item $h^{+}$  -   density of the holes.
\item $Tl^{0}_{e^{-}}$ -    density of trapped electrons at $Tl^0$ levels.  
\item $Tl^{++}_{h^{+}}$ -  density of trapped holes at $Tl^{++}$ levels.
\item  $Tl^{+*}_{popul}$  -    population of  $(Tl^{+})^{*}$ excited  levels.
\end{enumerate}

The parameters are also shown below:
\begin{enumerate}[(1)]
\item $\tau_{hop}$ - the  lifetime at  the STE level,  combined with the  time an  exciton needs, via hopping,   to reach the Tl excited level.
\item $\tau_{diss}$ - the  time an STE needs to decay to  electron and hole via  multiphonon dissociation.
\item $\tau^{0}_{trap}$ -  the lifetime of an  electron before it is trapped at $Tl^0$.
\item $\tau^{++}_{trap}$ -  the lifetime of a   hole before it is trapped at $Tl^{++}$. 
\item $\tau_{aftgl}$ -  the time an electron trapped at the $Tl^0$ level needs to reach and recombine with a hole trapped at $Tl^{++}$ via  thermoactivation. 
\item $\tau_{diff}$ -   the time a hole needs, after detrapping from the $Tl^{++}$ level, to reach and recombine with an electron  trapped at $Tl^0$ via diffusion.
\item $\tau_{scint} $ - the lifetime at  excited Tl level.
\item $\tau^{++}_{dtrap}$ - the  lifetime of  a trapped hole at the $Tl^{++}$ level.
\item$\tau^{0}_{dtrap}$ - the  lifetime of a trapped  electron at  the $Tl^0$ level.
\end{enumerate}

The most important parameters we used are well known: the lifetime of the trapped hole is  approximately $0.350$ $\mu$sec \cite{18}, and the lifetime of the $(Tl^{+})^*$ excited levels  is approximately 0.134 $\mu$sec \cite{5}. Less well known are the other parameters such as the multiphonon dissociation time, phonon-assisted hopping time, etc., which we varied. But provided these parameters are kept within reasonable ranges, the overall results are insensitive to their values. The values we used for $T= - 20^{0}$C and for $T=+50^{0}C$ are shown in Table I.

\begin{table}[!h]
\caption{Parameters Used in  Simulations  (nsec)} 
\begin{ruledtabular}
\begin{tabular}{lccccccccr}
$T[C]$ &  $\tau_{hop}$    &   $\tau_{diss}$  &  $\tau^{0}_{trap}$ &  $\tau^{0}_{dtrap}$  &  $\tau^{++}_{trap}$ & $\tau^{++}_{dtrap}$  &  $\tau_{diff}$ & $ \tau_{scin}$ & $\tau_{aftgl}$ \\
\hline
$-20^0 $   & $90$ & $70$ & $4$ & $200$  & $1$ & $350$ & $1$ & $134$ & $2.10^5$ \\
\hline
$+50^0$   & $4.2$  & $ 60 $ & $50$ & $20$  & $50$ & $350$ & $10^{-2}$ & $134$ & $10^3$ \\[1ex]
\end{tabular}
\end{ruledtabular}
\label{table:1} 
\end{table}
The obtained solutions fit our experimental results very well, as can be seen in Fig 4. This figure clearly shows the overall nonexponential form of both  the experimental results and  the model.

\begin{figure}[!h]
\begin{center}
\includegraphics[width=\columnwidth]{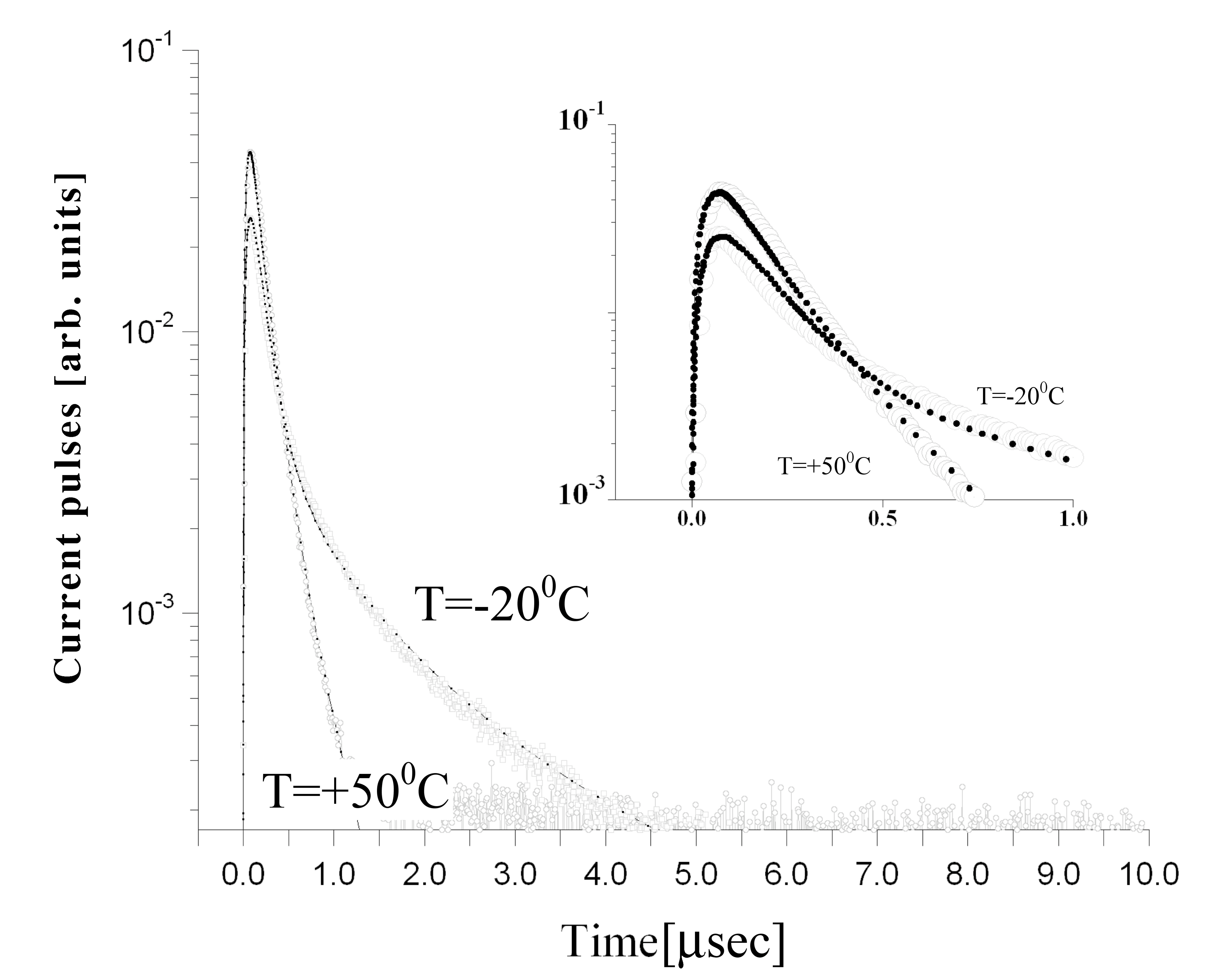}
\end{center}   
\caption{Two of the  experimentally measured current pulses normalized by area for NaI(Tl) at $-20^0C$ (upper curve) and $+50^0C$ (lower curve) are shown in grey dots. The numerical solutions of the nonlinear rate-equation model are shown in black dots. The inset shows the same pulses on a 1 $\mu$sec scale.}
\label{fig:4}
\end{figure} 
\section{CONCLUSIONS}

We have presented a model of two competing transport processes that explains the existence of two temporal components in the scintillation of NaI(Tl). While the linear temperature dependence of the total light output is due to the presence of a 
nonradiative decay channel from the activator level, a much stronger temperature  dependence in NaI(Tl) arises because of  the temperature-activated type dependence of the ratio between the fast and slow components. These two nonexponential components correspond to two distinct mechanisms of transport to the activator levels. One of the predictions made by the model is that the differential light output as a function of the energy of the ionizing particle will depend on the temperature. Furthermore, the temperature dependence of the shape of the light pulses, for a given initial energy, will depend on the level of doping because it arises form the competition  between unimolecular and bimolecular processes. In a subsequent publication, we will consider these topics, and we will argue that the difference in the transition rates between NaI(Tl) and CsI(Tl) is determined by the differences between the available phonon density of states at the activation energy for these two materials. 

Acknowledgements: Work at Los Alamos was performed under the auspices of the US Department of Energy. We wish to thank Cal Moss and K. B. Blagoev for useful comments.


\begin{thebibliography}{}
\bibitem{1}M. Moszyski, M. Balcerzyk, W. Czarnacki, M. Kapusta, W. Klamra, P. Schotanus, A. Syntfeld, M. Szawlowski, Nucl. Instrum. Methods, A537, 357-362 (2005). 
\bibitem{2}S.E. Derenzo, M.J. Weber, W.E. Bourret-Courchesne, M.K. Klintenberg,  Nucl. Instrum. Methods, 505 (1-2): 111-117 (2003). 
\bibitem{3}J.Biks,"The theory and practice of scintillation", Pergamon Press, (1964). 
\bibitem{4}Glenn F. Knoll, "Radiation detection and measurement", John WileySons (1999).  
\bibitem{6}L. M. Bollinger and G. E. Thomas, The Review of Scientifc Instruments, Vol. 32, (1961). 
\bibitem{5}J.S. Schweitzer	et al, IEEE Transaction on Nuclear Science, Vol. NS-30, No1 (1983).
\bibitem{7}Marku Koskelo et al, presented at INMM 46th annual meeting , Phoenix, Arizona (2005).
\bibitem{8}G. Pausch et al, presented at IEEE NSS-MIC Conference, Puerto Rico, Oct. (2005).
\bibitem{23}N. F. Mott and Gurney, Trans. Faraday Soc. Vol. 35,69,(1939).
\bibitem{9} K. D. Ianakiev, M.C. Browne, J. Audia, W. Hsue, "Apparatus and method for temperature correction and expanded count rate of inorganic scintillation detectors",US Patent  7,081,626 July 2005. 
\bibitem{11} K. D. Ianakiev et al, presented at IEEE NSS-MIC Conference, Puerto Rico, Oct. (2005), LA-UR 05-6822, and LA-UR 05-3409. 
\bibitem{12} K. D. Ianakiev et al. submitted in Nuclear Instrum. Methods.
\bibitem{10} B. Alexandrov et al, proceedings of the  INMM 46th Annual meeting, Phoenix, Arizona, July, (2005), LA-UR 05-4933.
\bibitem{24}Photonis, "Photomultiplier Tubes Principles and Applications", Brive, France (2002).
\bibitem{22} J.Valentine et all, Nuclear Instruments and Methods in Phys.Research A325 147-157C.L.(1993). 
\bibitem{13}  W.J. Van Sciever, 1955 High Energy Physics Rpt, No 38 Stanford University (1955). 
\bibitem{14} R. B. Murray and A. Meyer, Phys. Rev. vol. 122,  815 (1961). 
\bibitem{19} M. Tomura and Y. Kaifu, J. Phys. Soc. Of Japan, vol. 15., 1295 (1960) .
\bibitem{20} H. Nishimura and S. Nagata, Journal of Luminescence 4041, 429-430, (1988).
\bibitem{15} R.Gwin and R.B. Murray, Phys. Rev. vol. 131, 501, I (1963).
\bibitem{16} R.Gwin and R.B. Murray, Phys. Rev. vol. 131, 508 , II (1963).
\bibitem{17} R. B. Murray, IEEE Transaction on Nuclear Science, Vol. NS-22, (1975). 
\bibitem{18} H. B. Dietrich, A.E. Purdue, R. B. Murray and R.T. Williams, Phys. Rev. B vol. 8, 5894 (1973).
\bibitem{21} B. Goodman and O.S. Oen, J. Phys. Chem. Solids Vol. 8,. 291 -294 (1959). 
\end{thebibliography}
\end{document}